\documentclass[11pt]{article}
\usepackage[textwidth=15.2cm,textheight=22cm]{geometry}
\usepackage{amsmath,amssymb}
\usepackage{latexsym}
\usepackage{multicol}
\usepackage{graphicx}
\usepackage{bm}
\usepackage{varwidth}
\usepackage{pifont}
\usepackage{cancel}
\usepackage{xcolor}
\usepackage{tikz-feynman,contour} 
\usepackage{caption}
\usepackage{cite}
\usepackage{tikz}
\usetikzlibrary{shapes.geometric}
\usepackage{pifont}

\allowdisplaybreaks[1]

\newcommand{\del}{\partial}
\newcommand{\be}{\begin{equation}}
\newcommand{\ee}{\end{equation}}
\newcommand{\ba}{\begin{eqnarray}}
\newcommand{\ea}{\end{eqnarray}}
\newcommand{\bdm}{\begin{displaymath}}
\newcommand{\edm}{\end{displaymath}}

\newcommand{\rom}[1]{\uppercase\expandafter{\romannumeral #1\relax}}

\def\ba{\bar A}

\def\beq{\begin{equation}}
\def\eeq{\end{equation}}

\newcommand{\nn}{\nonumber}

\newcommand{\ndt}{\noindent}

\def\bea{\begin{eqnarray}}
\def\eea{\end{eqnarray}}
\def\beas{\begin{eqnarray*}}
\def\eeas{\end{eqnarray*}}
\def\sla{\raise.15ex\hbox{$/$}\kern-.57em}

\def\spa#1.#2{\left\langle#1\,#2\right\rangle}
\def\spb#1.#2{\left[#1\,#2\right]}

\begin{document}
\begin{titlepage}
\begin{flushright}    
{\small $\,$}
\end{flushright}
\vskip 1cm
\centerline{\Large{\bf{Deriving interaction vertices in higher derivative theories}}}
\vskip 0.7 cm
\centerline{Sudarshan Ananth, Nipun Bhave, Chetan Pandey and Saurabh Pant}
\vskip 0.3cm
\centerline{\it {Indian Institute of Science Education and Research}}
\centerline{\it {Pune 411008, India}}
\vskip 1.5cm
\centerline{\bf {Abstract}}
\vskip .2cm
\ndt We derive cubic interaction vertices for a class of higher-derivative theories involving three arbitrary integer spin fields. This derivation uses the requirement of closure of the Poincar\'e algebra in four-dimensional flat spacetime. We find two varieties of permitted structures at the cubic level and eliminate one variety, which is proportional to the equations of motion, using suitable field redefinitions. We then consider soft theorems for field theories with these higher-derivative interactions and construct amplitudes in these theories using the inverse-soft approach.
\vskip .5cm
\ndt 
\vfill
\end{titlepage}

\section{Introduction}
\ndt The study of scattering amplitudes has revealed surprising simplicity in the mathematical structures underlying Yang-Mills theories. The past few decades have seen impressive progress in our understanding of amplitudes and our efficiency in computing them. Amplitudes exhibit a number of interesting properties and satisfy a variety of relations (KLT, BCJ, color-kinematics and so on). The light-cone gauge offers a not-so-mainstream perspective on scattering amplitudes: with both locality and covariance being non-manifest, this gauge eliminates unphysical degrees of freedom at the cost of making computations more technical. Importantly, spurious degrees of freedom and redundancies do not obscure symmetries in the theory - symmetries often being key to the search for simplicity (the compact spinor helicity variables also emerge naturally in this gauge). 
\vskip 0.25cm
\ndt  The classic paper~\cite{BBB1} presented the `derivation' of consistent cubic interaction vertices using just two ingredients: physical fields (unphysical degrees of freedom having been eliminated) and the Poincar\' e algebra (which must close). However, this study did not include higher-derivative corrections, which often appear in effective actions (and serve as potential counter terms in loop amplitudes). Such terms were precluded by the choice of length dimension $L^{\lambda-1}$ for the coupling constant ($\lambda$ being the helicity of the fields). There has been considerable work on constructing consistent interaction vertices in the light front approach using the Fock-space method \cite{Bengtsson:1986kh,Metsaev:1991nb,Metsaev:1991mt,Metsaev:1993ap,Metsaev:2005ar,Bengtsson:2016hss}  and in momentum space \cite{Ponomarev:2016cwi,Ponomarev:2022vjb}. 
\vskip 0.25cm
\ndt This paper expands the framework of~\cite{BBB1} to include, beyond the usual structures, higher-derivative terms. The consequences of these terms and their implications for amplitude structures - which have close ties to the light-cone formalism~\cite{Ananth:2012un, Akshay:2014qea} - are examined. The inverse-soft method~\cite{Arkani-Hamed:2009ljj,Boucher-Veronneau:2011rwd,Nandan:2012rk} is then used to build higher-point amplitudes.
\vskip 0.25cm
\ndt Since the light-cone formalism is not covariant, Lorentz invariance needs to be verified. The key idea is to convert this `task' into a tool, using it to constrain and then determine the Hamiltonian entirely. This allows us to construct cubic interaction vertices for a class of higher-derivative theories. This approach is also generalized to higher-point vertices and as an example, the quartic vertex is constructed for the simplest possible higher derivative operator. We also invoke symmetry arguments to explain the permissible structures for $n$-point interaction vertices. 
\vskip 0.25cm

\ndt Scattering amplitudes for a large class of higher derivative operators have been studied in the literature previously~\cite{Dixon:2004za,Broedel:2012rc, Cohen:2010mi,Cheung:2015cba,He:2016iqi} using methods like CSW, BCFW, CHY and color-kinematic duality. These operators are not generally constructible, because of the potential boundary term. However, there have been some attempts to recursively construct a class of amplitudes for higher derivative operators using BCFW or all-line shift method \cite{Cohen:2010mi,Cheung:2015cba}. But in general, higher derivative theories are not constructible. In this paper, we attempt to recursively construct scattering amplitude for higher derivative theories using inputs from soft theorems. We use the inverse soft method, a complementary technique to those mentioned above, to derive higher-point tree-level amplitudes \cite{Boucher-Veronneau:2011rwd,Nandan:2012rk}. In this approach, lower-point amplitudes are multiplied by a universal soft factor with appropriate legs shifted. This method is equivalent to BCFW recursion relations. In fact for MHV amplitudes, the inverse soft method is much simpler than the other known recursion relation methods. This method can only be used to construct amplitudes if there is no pole at infinity. Starting with the derived cubic interaction vertex as a seed amplitude, we construct MHV amplitudes for a class of higher derivative theories. We then extend our construction to higher-point NMHV amplitudes by starting with the known seed amplitudes, using the inverse soft technique to recursively contruct higher-point NMHV amplitudes.

\vskip 1cm

\section{Construction of cubic interaction vertices for higher-dimensional operators }
\ndt We define light-cone co-ordinates in $(-,+,+,+)$ Minkowski space-time as
\begin{eqnarray}
x^{\pm}=\frac{x^{0}\pm x^{3}}{\sqrt{2}} \;,\qquad
x = \frac{x^{1}+ix^{2}}{\sqrt{2}} \;,\qquad\bar{x}= x^*\,,
\end{eqnarray}
\ndt with  $\partial_\pm, \,\bar\partial,\,\partial$\, being the corresponding derivatives and the operator $\frac{1}{\partial^+}$ defined following the prescription in~\cite{SM}. $x^+$ is chosen as the time coordinate so $p^-$ is the light-cone Hamiltonian.\\

\ndt The Poincar\'e algebra in these coordinates is realized on the two physical degrees of freedom $\phi$ and $\bar{\phi}$. The Poincar\'e generators split into two types: kinematical $\mathbb{K}$ which do not involve the time derivative $\del_+$ and dynamical $\mathbb{D}$ which do - and hence pick up non-linear contributions in the interacting theory \cite{BBB1}. The generators are
\bea
&&\mathbb{K}: \;\;\{ p,\, \bar{p}, \,p^+,\, j, \,j^+, \,\bar{j}^+,\, j^{+-} \} \,, \nn\\
&&\mathbb{D}: \;\; \{ p^- \equiv H,\, j^-,\, \bar{j}^- \} \, .
\eea
The algebraic structures are 
\bea
[\mathbb{K}\,,\mathbb{K}] = \mathbb{K}\,, \;\;\;\;\;\; [\mathbb{K}\,,\mathbb{D} ] = \mathbb{D}\,, \;\;\;\;\;\; [\mathbb{K}, \mathbb{D}] = \mathbb{K}\,,\;\;\;\;\;\; [\mathbb{D}, \mathbb{D} ] = 0\,.
\eea
\ndt Here, we review key features of this formalism and refer the reader to appendix A for additional details.  \\

\ndt The Hamiltonian for the free field theory is
\begin{equation}
\label{free}
H =-\int d^3x\,{\bar\phi}_i\,\partial\bar\partial\,\phi_i\, = \int d^{3}x\,\partial^+{\bar\phi}_i\,\delta_{p^-}\phi_i\ ,
\end{equation}
with $\phi_i$ referring to a field of helicity $\lambda_i$ with $i\in \mathbb{Z}^{+}$. Upon switching `on' interactions, the $\delta_{p^-}$ operator picks up corrections, order by order, in the coupling constant $\alpha$.

\vskip 0.3cm
\ndt Reference~\cite{BBB1} focused on the case of interactions between fields, all having  helicity $\lambda$ fields with $\alpha$ having dimensions of $L^{\lambda-1}$. 
\vskip 0.3cm
\ndt In this paper, we consider instead the following two - most general - ansatze for cubic interactions (based on dimensional analysis and helicity counting)
\bea
\label{h1}
\text{Type-1\,:}\;\;\;\;\; \delta^{'\alpha}_{{p}^{-}} \,\phi_1  = \alpha \,A\, \del^{+\,\mu} \left[\, \bar{\del}^a\del^c \del^{+\, \rho} \bar{\phi}_2\, \bar{\del}^b\del^d \del^{+\, \sigma} \bar{\phi}_3\,\right]\,,
\eea
\bea
\label{h2}
\text{Type-2\,:}\;\;\;\;\;\delta^{''\alpha}_{{p}^{-}} \phi_1  =\alpha\, C \,\del^{+\,\mu} \left[\, \bar{\del}^a\del^c \del^{+\, \rho} {\phi}_2\, \bar{\del}^b\del^d \del^{+\, \sigma} {\phi}_3\,\right] \,,
\eea
\ndt  where $\mu, \rho,\sigma, a,b$ are integers and $A$ and $C$ are numerical factors.\\
\vskip 0.1cm
\ndt The key departure from~\cite{BBB1} for both types of ansatze being that the dimension of the coupling constant is $[\alpha] =L^{\lambda_2 +  \lambda_3  +  \lambda_1- 1}$. This choice will permit us to derive cubic interaction vertices through the algebra-closure method in a new class of theories - higher derivative theories, formulated in the light-cone gauge~\footnote{For a review of the Ostrogradskian constraints associated with higher-derivative theories, please see~\cite{RPW}.}.
\vskip 0.3cm

\subsection{Type-1 cubic interaction vertices}
\ndt We first start with $\delta^{'\alpha}_{{p}^{-}} \phi_1$ and use the commutation relations and dimensional analysis to find the unknown parameters. The commutators 
\bea
&&[\,\delta_j \,, \delta^{'\alpha}_{{p}^{-}}\,]\phi_1 = 0 \,, \nn\\
&& [\,\delta_{j^{+-}} \,, \delta^{'\alpha}_{{p}^{-}}\,]\phi_1 = -\delta^{'\alpha}_{{p}^{-}}\phi_1\,,
\eea
\ndt imposes the following conditions on our ansatz 
\bea
\label{C01}
&&(c+d)-(a+b)  =  \lambda_2 +  \lambda_3  +  \lambda_1 \,, \nn\\
&& \mu + \rho + \sigma = -1 \,.
\eea	
\ndt Let $ \lambda =\lambda_2 +  \lambda_3  +  \lambda_1$ so the first equation of (\ref{C01}) reads $(c+d)-(a+b)  = \lambda$. The dimensional analysis of (\ref{h1}) gives us the following relation
\bea
\label{C02}
(a+b) +(c+d) = \lambda  \,.
\eea
\ndt Adding the first equation of (\ref{C01}) and (\ref{C02}), we get
\bea
\label{d1}
&&c+d = \lambda\,, \\
&&a+b = 0\,.
\eea
\ndt As $a,b,c,d > 0$, this implies that $a=b=0$. Therefore, mixed derivative terms are not allowed for type-1 vertices (\ref{h1}). As $c+d=\lambda$, there are $\lambda +1 $ possible values for a pair $(c,d)$. We rewrite the ansatz (\ref{h1}) as a sum of these $\lambda +1 $ terms
\bea
\label{h11}
\delta^{'\alpha}_{{p}^{-}} \phi_1  = \alpha\,\sum_{n=0}^{\lambda}\, A_n \del^{+\,\mu_n} \,\left[\, \del^{(\lambda -n)} \del^{+\, \rho_n} \bar{\phi}_2\, \del^n \del^{+\, \sigma_n} \bar{\phi}_3\,\right]  \,.
\eea

\ndt The next commutator $[\,\delta_{\bar{j}^+} \,, \delta^{'\alpha}_{{p}^{-}}\,]\phi_1 = 0$ gives the following condition
\bea
&&\sum_{n=0}^{\lambda}A_n \bigg\{  \,(\lambda -n)\,\del^{+\,\mu_n} \,\left[\, \del^{(\lambda -n-1)} \del^{+\, (\rho_n + 1)} \bar{\phi}_2\, \del^n \del^{+\, \sigma_n} \bar{\phi}_3\,\right]  \nn\\
&& \hspace{1.8 cm}+ \, n \,\del^{+\,\mu_n} \,\left[\, \del^{(\lambda -n)} \del^{+\, \rho_n } \bar{\phi}_2\, \del^{(n-1)} \del^{+\, (\sigma_n+1)} \bar{\phi}_3\,\right] \bigg\} =0 \,.
\eea
\ndt The above condition is satisfied if the coefficients obey the following recursion relations
\bea
&&A_{n+1} = - \frac{(\lambda - n)}{(n+1)} A_n = (-1)^{n+1} {\lambda \choose n+1}A_0 \,,  \\
&& \rho_{n+1} = \rho_n+1 \,,\;\;\; \sigma_{n+1} = \sigma_n-1 \,,\;\;\; \mu_{n+1} = \mu_n\,.
\eea
\vskip 0.1 cm
\ndt To determine the exact values of $\rho,\mu,$ and $\sigma$, we need the dynamical commutators $[\,\delta_{{j}^{-}} \,, \delta^{'}_{{p}^{-}}\,]^{\alpha}\,\phi_1 = 0\,$ and $[\,\delta_{\bar{j}^{-}} \,, \delta^{'}_{{p}^{-}}\,]^{\alpha}\,\phi_1 = 0\,.$ The boost generators $j^-$ and $\bar{j}^{-}$ also get corrected when interactions are turned on and are of the form
\bea
\delta_{{j}^{-}}\phi = -{x} \,\delta^{'\alpha}_{{p}^{-}}\phi + \delta_{{s}}^{\alpha}\phi \,,\hspace{2.5 cm}\delta_{\bar{j}^{-}}\phi = -\bar{x} \,\delta^{'\alpha}_{{p}^{-}}\phi + \delta_{\bar{s}}^{\alpha}\phi\,.
\eea
\vskip 0.1 cm
\ndt The boost generators are determined if we know the spin parts $\delta_{{s}}^{\alpha}\phi\,, \,\delta_{\bar{s}}^{\alpha}\phi$. For type-1 vertices they are structurally of the form 
\bea
\delta_{\bar{s}}^{\alpha}\phi \sim \del^{\lambda-1} \bar{\phi}\,\bar{\phi} \,,\hspace{2 cm} \delta_{{s}}^{\alpha}\bar{\phi} \sim \bar{\del}^{\lambda-1} \phi \phi \,. 
\eea
Due to helicity, the transformations $\delta_{{s}}^{\alpha}\phi$ and $\delta_{\bar{s}}^{\alpha}\bar{\phi}$ do not exist. We now compute $ [\,\delta_{j^{-}} \,, \delta'_{{p}^{-}}\,]^{\alpha}\,\phi_1 = 0$ to obtain
\bea
&&\sum_{n=0}^{\lambda}A_n \bigg\{  \,(\mu_n+\lambda_1+1)\,\frac{\del}{\del^+}\,\del^{+\,\mu_n} \,\left[\, \del^{(\lambda -n)} \del^{+\, \rho_n } \bar{\phi}_2\, \del^n \del^{+\, \sigma_n} \bar{\phi}_3\,\right]  \nn\\
&& \hspace{1.8 cm}+ \, (\rho_n + \lambda_2) \,\del^{+\,\mu_n} \,\left[\, \del^{(\lambda -n+1)} \del^{+\, (\rho_n-1) } \bar{\phi}_2\, \del^{n} \del^{+\, \sigma_n} \bar{\phi}_3\,\right] \nn\\
&& \hspace{1.8 cm}+ \, (\sigma_n + \lambda_3) \,\del^{+\,\mu_n} \,\left[\, \del^{(\lambda -n)} \del^{+\, \rho_n } \bar{\phi}_2\, \del^{(n+1)} \del^{+\, (\sigma_n-1)} \bar{\phi}_3\,\right] \bigg\} =0\,.
\eea
\vskip 0.1 cm
\ndt The solution of the above recursion relation for $\rho, \sigma$ and $\mu$ subject to the boundary conditions $\sigma_{n=\lambda} = -\lambda_3 \,,\; \rho_{n=0} = -\lambda_2 \,$ is 
\bea
\rho_n = n - \lambda_2 \,, \;\;\;\;\; \sigma_n = \lambda - \lambda_3- n \,,\;\;\;\; \mu_n = -1 - \lambda_1\,.
\eea
\vskip 0.2 cm
\ndt Plugging the values of $\rho, \sigma$ and $\mu$ in our ansatz, we find
\bea
\delta^{'\alpha}_{{p}^{-}} \phi_1  = \sum_{n=0}^{\lambda}\, (-1)^n {\lambda \choose n}\, \del^{+\,(-1- \lambda_1) } \,\left[\, \del^{(\lambda -n)} \,\del^{+\, (n-\lambda_2)} \,\bar{\phi}_2\, \del^n \,\del^{+\, (\lambda - \lambda_3- n)} \,\bar{\phi}_3\,\right]  \,.
\eea
Since
\bea
H = \int d^3x \; \del^+ \bar{\phi}_1 \,\delta^{'\alpha}_{{p}^{-}} \phi_1 \,,
\eea
the interaction Hamiltonian is 
\bea
\label{H11}
H^{\alpha} = \alpha\int d^3x \; \,\sum_{n=0}^{\lambda}\, (-1)^n {\lambda \choose n}\, \frac{1}{\del^{+\,\lambda_1}} \bar{\phi}_1 \,\left[\, \del^{(\lambda -n)} \,\del^{+\, (n-\lambda_2)} \,\bar{\phi}_2\, \del^n \,\del^{+\, (\lambda - \lambda_3- n)} \,\bar{\phi}_3\,\right] + c.c. \,.
\eea
\\
As is well known, for odd $\lambda$, non-trivial cubic vertices require the introduction of an antisymmetric structure constant $f^{abc}$. 
\vskip 0.4 cm
\subsubsection{Amplitude structures}
\ndt In momentum space, the cubic vertices (\ref{H11}) have the following structure (with measure and constants suppressed)
\bea
H^{\alpha} &=&   \,\sum_{n=0}^{\lambda}\, (-1)^n {\lambda \choose n}\,  \frac{1}{p^{+\,\lambda_1}}  \,\left[\, k^{(\lambda -n)} \,k^{+\, (n-\lambda_2)} \, l^n \,l^{+\, (\lambda - \lambda_3- n)} \,\right] \bar{\phi}_1(p)\,\bar{\phi}_2(k)\,\bar{\phi}_3(l) + c.c. \,,\nn\\
&=& \, \frac{k^{+\,(\lambda_1 + \lambda_3)}\,l^{+\,(\lambda_1 + \lambda_2)}}{p^{+\,\lambda_1}}\,\sum_{n=0}^{\lambda}\, (-1)^n {\lambda \choose n}\,    \,\left[\, \left(\frac{k}{k^{+}}\right)^{(\lambda -n)}\; \left(\frac{l}{l^+} \right)^n \right] \bar{\phi}_1(p)\,\bar{\phi}_2(k)\,\bar{\phi}_3(l) + c.c. \,,\nn\\
&=& \frac{\left(kl^+ - lk^+ \right)^{\lambda}}{p^{+\,\lambda_1}\,k^{+\,\lambda_2}\,l^{+\,\lambda_3}} \;\bar{\phi}_1(p)\,\bar{\phi}_2(k)\,\bar{\phi}_3(l) + c.c.  \,.
\eea
\ndt  The off-shell spinor products in this language are
\bea
\langle kl\rangle \equiv \sqrt{2}\,\frac{(kl^{+}-lk^{+})}{\sqrt{k^{+}l^{+}}},\hspace{1cm}  [ kl] \equiv \sqrt{2}\,	\frac{(\bar kl^{+}-\bar lk^{+})}{\sqrt{k^{+}l^{+}}}\ .
\eea
\ndt In terms of spinor helicity variables~\cite{Ananth:2012un}, the vertex reads
\bea
\label{V1}
V^{\alpha}(p,k,l) =\frac{1}{\sqrt{2^{\lambda}}}\;\langle pk \rangle ^{\lambda_1 + \lambda_2 - \lambda_3}\, \langle kl \rangle ^{\lambda_2 + \lambda_3 - \lambda_1}\, \langle lp \rangle ^{\lambda_3 + \lambda_1 - \lambda_2} + c.c. \,.
\eea
\ndt This is consistent with the general result for three-point amplitudes derived in \cite{Benincasa:2007xk, EH } using S-matrix arguments, and little group scaling and in~\cite{Bengtsson:1986kh,Metsaev:1991nb} using a Fock-space approach.

\vskip 0.3 cm

\subsection{Type-2 cubic interaction vertices}
\label{t2}
\ndt We start with (\ref{h2})
\bea
\label{h2222}
\delta^{''\alpha}_{{p}^{-}} \phi_1  =\alpha\, C \,\del^{+\,\mu} \left[\, \bar{\del}^a\del^c \del^{+\, \rho} {\phi}_2\, \bar{\del}^b\del^d \del^{+\, \sigma} {\phi}_3\,\right]  \,,
\eea
\ndt and compute the commutators 
\bea
&&[\,\delta_j \,, \delta^{''\alpha}_{{p}^{-}}\,]\phi_1 = 0 \,, \nn\\
&& [\,\delta_{j^{+-}} \,, \delta^{''\alpha}_{{p}^{-}}\,]\phi_1 = -\delta^{''\alpha}_{{p}^{-}}\phi_1\,,
\eea
\ndt to arrive at the following conditions
\bea
\label{C1}
&&(a+b) -(c+d) =  \lambda_2 +  \lambda_3  -  \lambda_1  \,, \\
&& \mu + \rho + \sigma = -1 \,.
\eea	
We also have, from dimensional analysis,
\bea
\label{C2}
(a+b) +(c+d) = \lambda_2 +  \lambda_3  +  \lambda_1 \,.
\eea
\ndt Adding (\ref{C1}) and (\ref{C2})
\bea
&&a+b = \lambda_2 +  \lambda_3 \,,  \\
&&c+d = \lambda_1 \,.
\eea
\ndt Note that if $\lambda_1=0$ then (\ref{h2222}) becomes a type-1 vertex. We now encounter a double sum as opposed to the single sum in (\ref {h11}). We need an index $n$ associated with the $\lambda_2 +  \lambda_3 +1$ possible values the pair $(a,b)$ can take and an index $m$ for the $\lambda_1+1$ values that the pair $(c,d)$ run over. We rewrite our ansatz (\ref{h2}) as the double sum
\bea
\label{h22}
\delta^{''\alpha}_{{p}^{-}} \phi_1  = \alpha\,\sum_{n=0}^{\lambda_2 +  \lambda_3 }\,\sum_{m=0}^{\lambda_1} C_{n,m} \,\del^{+\,\mu_{n,m}} \left[\, \bar{\del}^{(\lambda_2 +  \lambda_3 -n)}\,\del^{(\lambda_1-m)} \,\del^{+\, \rho_{n,m}}\, {\phi}_2\, \bar{\del}^n\,\del^m \,\del^{+\, \sigma_{n,m}}\, {\phi}_3\,\right]  \,.
\eea
\ndt The detailed calculation for this variety of vertex is presented in appendix B. \\
\vskip 3 mm
\ndt We find, for the type-2 vertex, 
\bea
\label{h221}
\delta^{''\,\alpha}_{{p}^{-}} \phi_1 & =& \alpha\,\sum_{n=0}^{\lambda_2 +  \lambda_3 }\,\sum_{m=0}^{\lambda_1}\, (-1)^{(n+m)}\,{\lambda_2+\lambda_3 \choose n}\,{\lambda_1 \choose m} \nn\\
&& \bigg\{\,\del^{+\,{\mu} }  \left[\, \bar{\del}^{(\lambda_2 +  \lambda_3 -n)}\,\del^{(\,\lambda_1-m)} \,\del^{+\, {\left(\,n+m+u \right)}} {\phi}_2\, \bar{\del}^n\,\del^m \,\del^{+\, {\left(v- n-m\,\right)}} {\phi}_3\,\right] \bigg\} \,, 
\eea
with $ n+m + u= \rho_{n,m}\,,\, v-n-m=\sigma_{n,m}$, $\mu=\mu_{n,m}$ where u, v are functions of the $\lambda_i$.\\

\ndt Poincar\'e invariance is insufficient to uniquely fix the form of cubic interaction vertices of type-2. This is because the helicity constraints permit a non-zero term in the spin part of the boost generator, ie. $\delta_{\bar{s}} \phi$ (see appendix B), which is disallowed for type-1 vertices. Physically this non-zero spin part can be thought of as a loop correction to the usual spin transformation (equation 3.26 of \cite{BBB1}). This was noted in \cite{Bengtsson:2012dw} where three-point counterterms were constructed for gravity in the light-front formalism. In that work, it was suggested that additional symmetry is necessary to uniquely determine the exact form of the counterterms. In the case of gravity, the residual gauge symmetry was used to fix the exact form of the three-point counterterm. To determine the type-2 vertex uniquely, an analog of residual gauge symmetry is likely to be necessary. \\ 

\ndt Since the type-2 vertex contains both kinds of derivatives, it can be shown to be proportional to the free equations of motion  \cite{Bengtsson:1986kh,Bengtsson:2012jm} {\it {at cubic order}}. We rewrite (\ref{h221}) as
\bea
\delta^{''\,\alpha}_{{p}^{-}} \phi_1 & =& \alpha\,\sum_{n=0}^{\lambda_2 +  \lambda_3 }\,\sum_{m=0}^{\lambda_1}\, (-1)^{(n+m)}\,{\lambda_2+\lambda_3 \choose n}\,{\lambda_1 \choose m}\nn\\
&&  \bigg\{\,\del^{+\,{\mu} }  \left[\, \bar{\del}^{(\lambda_2 +  \lambda_3 -n)}\,\del^{(\,\lambda_1-m)} \,\del^{+\, {\left(\,n+m+u \right)}} {\phi}_2\, \bar{\del}^n\,\del^m \,\del^{+\, {\left(v- n-m\,\right)}} {\phi}_3\,\right] \bigg\}\,, \nn\\
&=& -\frac{1}{2}\,\alpha\, {\square}\sum_{n=0}^{\lambda_2 +  \lambda_3 -1}\,\sum_{m=0}^{\lambda_1-1}\, (-1)^{(n+m)}\,{\lambda_2+\lambda_3 -1\choose n}\,{\lambda_1-1 \choose m}  \nn\\
&& \hspace{-1 cm}  \bigg\{\,\del^{+\,{\mu} }  \left[\, \bar{\del}^{(\lambda_2 +  \lambda_3 -n-1)}\,\del^{(\,\lambda_1-m-1)} \,\del^{+\, {\left(\,n+m+u+1 \right)}} {\phi}_2\, \bar{\del}^n\,\del^m \,\del^{+\, {\left(v- n-m-1\,\right)}} {\phi}_3\,\right] \bigg\}\,,
\eea
\ndt where we have used the equation of motion $\del^- \phi = \frac{\del \bar{\del}}{\del^+} \phi + O(\phi^2)$. So, type-2 cubic vertices are proportional to the free equations of motion. Therefore, for this class of higher derivative theories, with only type-1 cubic vertices (\ref{H11}), the  Hamiltonian is
\bea
\label{Hf1}
&&H^{\alpha} = \int d^3x \; \del^+ \bar{\phi}_1 \,\delta^{\alpha}_{{p}^{-}} \phi_1 = \del^+ \bar{\phi}_1 \,\delta^{'\alpha}_{{p}^{-}} \phi_1 \,, \\
 &&= \alpha\int d^3x \; \,\sum_{n=0}^{\lambda}\, (-1)^n {\lambda \choose n}\, \frac{1}{\del^{+\,\lambda_1}} \bar{\phi}_1 \,\left[\, \del^{(\lambda -n)} \,\del^{+\, (n-\lambda_2)} \,\bar{\phi}_2\, \del^n \,\del^{+\, (\lambda - \lambda_3- n)} \,\bar{\phi}_3\,\right] + c.c. \,. \nn
\eea
\ndt For example, a $R^2$ type operator, based on dimensional analysis and helicity, can only produce a type-2 cubic interaction vertex \cite{Bengtsson:2012dw}. This being proportional to the equations of motion, may be removed by a suitable field redefinition (thus all $n$-point graviton amplitudes produced by the $R^2$ term vanish as expected~\cite{He:2016iqi}). \\

\ndt We can generalize this framework from cubic vertices to specific class of $n$-point interaction vertices as discussed below. We construct the simplest possible quartic vertex as a specific example. The details are presented in the appendix C.

\subsection*{Comments on $n$-point interaction vertices in higher derivative theories}
\ndt In this section, we first deduce the structure of interaction vertices at higher orders purely from dimensional and kinematical constraints and then prove that all $n$-point vertices containing purely one type of transverse derivatives can be uniquely fixed by the Poincar\'e algebra. \\

\ndt We work here with a special class of higher derivative theories where $\lambda_i=\lambda$, and work out the structure of interaction vertices at higher orders purely from symmetry constraints. In a perturbative expansion, the dimension of the coupling for a $n$-point interaction vertex is 
\bea
[\alpha^{n-2}]=3\lambda-1+(n-3)(\lambda-1)\,.
\eea
\ndt where $\alpha$ is the $3-$point coupling. The $n$-point Hamiltonian is of the form $\phi^p\,\bar{\phi}^q$. We start with the ansatz
 \bea 
\label{HFF}
\delta_{{p}^{-}}^{\alpha^{n-2}} \phi&=&\alpha^{n-2}\,{\partial^{+}}^{\mu_0}\biggl \{[{\bar\partial}^{{a}_1} \,\partial^{{c}_1}\, {\partial^{+}}^{{\mu}_1}  \phi]\;\,[\,{\bar\partial}^{{a}_2} \,\partial^{{c}_2}\, {\partial^{+}}^{{\mu}_2}  \phi]\,........\,[{\bar\partial}^{{a}_p} \,\partial^{{c}_p}\, {\partial^{+}}^{{\mu}_p}\phi] \nn\\
&&\hspace{2 cm} \,\;[{\bar\partial}^{{a}_{p+1}} \,\partial^{{c}_{p+1}}\, {\partial^{+}}^{{\mu}_{p+1}}\bar\phi]\,.......\,
[{\bar\partial}^{{a}_{n-1}} \,\partial^{{c}_{n-1}}\,{\partial^{+}}^{{\mu}_{n-1}} \bar\phi] \biggr\} \,,\nn
\eea

\ndt where $p+q=n$ and the $a_i$, $ c_i$ are non-negative integers and the ${\mu}_i$ are integers. The commutator $[\delta_{j},\delta_{{p}^{-}}]$ yields
\bea
\label{delj}
\sum_{i=1}^{n-1}({a}_i- {c}_i) =  (p-q)\lambda \ .
\eea
Using the commutator $[\delta_{j^{+-}},\delta_{{p}^{-}}]$ gives $\sum_{i=0}^{n-1} {\mu}_i  = -1$. Dimensional analysis gives the following constraint
\bea
\label{dim}
\sum_{i=1}^{n-1} ({a}_i + {c}_i) = 6 + (\lambda - 2)(p + q )\ .
\eea

\ndt Using (\ref{delj}), (\ref{dim}) and the non-negativity of powers of transverse derivatives we obtain
\bea
\label{ineq1}
\sum_{i=1}^{n-1} {a}_i = (\lambda - 1)p - q  + 3 \geq 0\ ,\\
\label{ineq2}
\sum_{i=1}^{n-1} {c}_i = (\lambda - 1)q - p + 3 \geq 0\ .
\eea

\ndt For $\lambda=1$, we get $p,q\le 3$. At cubic order, note that the $(p=3,q=0)$ $A\,A\,A$ structure and $(p=0,q=3)$ $\bar{A}\,\bar{A}\,\bar{A}$ structure follow from this. At the next order, two new structures $A^3\,\bar{A}$ and $\bar{A}^3\,A$ are allowed as compared to the usual Yang-Mills quartic vertices. \\

\ndt We then consider $\lambda=2$, and use $q=n-p$ in (\ref{ineq1}) and (\ref{ineq2})
\bea
\frac{n-3}{2} \,\le\, p\,\le \, \frac{n+3}{2}\,.
\eea
\ndt The interaction vertex may be odd or even. For an odd point vertex, $n=2m+1$ where $m$ is a positive integer. This gives
\bea
\label{2-odd}
m-1 \,\le\, p\,\le \, m+2 \,.
\eea
\ndt Thus, for a cubic vertex where $m=1$, (\ref{2-odd}) allows helicity structures $h\,h\,h$ and $\bar{h}\,\bar{h}\,\bar{h}$. \\

\ndt For an even point vertex $n=2m$ we get
\bea
\label{2-even}
m-\frac{3}{2} \,\le\, p\,\le \,m+ \frac{3}{2} \,.
\eea
\ndt Since $p$ is an integer, the condition (\ref{2-even}) is
\bea
\label{2-evenf}
m-1 \,\le\, p\,\le \,m+1 \,.
\eea
\ndt For a quartic vertex where $m=2$, (\ref{2-evenf}) allows terms of helicity structure $h\,\bar{h}^3$ and $\bar{h}\,h^3$.  Thus, higher derivative operators produce new helicity configurations at each order. For example, in~\cite{Ananth:2022spf} it was shown for $n=6$, that only $h^3\,\bar{h}^3$ type of vertices occur. Here, additional vertices of type $h^2\,\bar{h}^4$ and $\bar{h}^2\,h^4$ appear at $n=6$. \\

\ndt We now prove that: \textit{all $n$- point vertices containing purely one type of transverse derivatives can be uniquely fixed by the Poincar\'e algebra.} \\

\ndt We start with an ansatz for the $n$-point interaction vertex (\ref{HFF}) and plug $a_i=0$. It reads
\bea 
\delta_{{p}^{-}}^{\alpha^{n-2}} \phi&=&\alpha^{n-2}\,{\partial^{+}}^{\mu_0}\biggl \{[ \,\partial^{{c}_1}\, {\partial^{+}}^{{\mu}_1}  \phi]\;\,[\, \,\partial^{{c}_2}\, {\partial^{+}}^{{\mu}_2}  \phi]\,........\,[ \,\partial^{{c}_p}\, {\partial^{+}}^{{\mu}_p}\phi] \nn\\
&&\hspace{2 cm} \,\;[ \,\partial^{{c}_{p+1}}\, {\partial^{+}}^{{\mu}_{p+1}}\bar\phi]\,.......\,
[ \,\partial^{{c}_{n-1}}\,{\partial^{+}}^{{\mu}_{n-1}} \bar\phi] \biggr\} \,,\nn
\eea

\ndt Consistency with the helicity generator $j$ requires
\bea
\sum_{i=1}^{n-1}c_i=(q-p)\lambda
\eea
\ndt The commutator with the generators $j^{+-}\;,\bar{j}^{\,+}$ determines the vertex upto the powers of $\partial^+$s. The exact powers of $\partial^+$s are fixed by the dynamical generators. We argue that the spin transformation appearing in the dynamical generator $j^{\,-}$ at this order is trivial and hence can be used to uniquely determine the vertex.\\

\ndt  The number of transverse derivatives in the spin transformation $\delta_s^{\alpha^{n-2}}\phi\,\sim\,\alpha^{n-2}\,\phi_1......\bar{\phi}_{n-1}$ (derivatives suppressed) must be one less than that in $ \delta_{{p}^{-}}^{\alpha^{n-2}} \phi$ due to its dimensionality. The dynamical generator $j^{\,-}$ has helicity $+1$. In order for the spin transformation $\delta_s^{\alpha^{n-2}}\phi$ to have helicity $+1$, the number of transverse derivatives in it must be one greater than that in $ \delta_{{p}^{-}}^{\alpha^{n-2}} \phi$. This proves that the spin transformation $\delta_s^{\alpha^{n-2}}\phi$ cannot be consistent with the helicity and dimensionality simultaneously and hence must vanish. This allows us to uniquely determine the vertex. Therefore, all $n$- point vertices containing purely one type of transverse derivatives can be uniquely fixed by the Poincar\'e algebra.\\

\subsection*{Vertices to Amplitudes : Motivation to use the inverse soft technique}

\ndt We would now like to compute amplitudes using the interaction vertices derived previously. In principle, one can construct the $n$-point tree-level amplitudes. However, the calculation becomes mathematically tedious as the number of Feynman diagrams and the number of terms in the diagram increases exponentially. For example, to calculate a 4-point tree-level amplitude, we need to sum over contributions from the exchange diagrams and the contact diagram. The contact term may be derived by the closure of Poincar\'e algebra in special cases (as discussed above) which in itself requires a lot of work. We therefore use a complementary technique, the inverse soft method, in the next section to derive higher-point tree-level amplitudes for the higher derivative theories (see appendix C). \\

\section{Inverse soft construction in higher derivative theories}

\ndt Soft factors in the light-cone gauge and their use in the construction of higher-point interaction vertices were the focus of~\cite{Ananth:2022cxo} (this included a review of the light-cone realization of many covariant results from~\cite{Arkani-Hamed:2009ljj,Boucher-Veronneau:2011rwd,Weinberg:1965nx}). In this section, we explore these methods in the context of the higher-derivative theories
considered in this paper. \\

\ndt The idea that higher-point tree level amplitudes can be constructed from the lower-point ones by using a multiplicative universal factor, associated with the emission of a soft boson was first presented in \cite{Arkani-Hamed:2009ljj} and then subsequently developed in \cite{Boucher-Veronneau:2011rwd,Nandan:2012rk}. It was shown in \cite{Boucher-Veronneau:2011rwd}, that inverse soft construction with BCFW  \cite{Britto:2004ap,Britto:2005fq} as a guide, can be used to construct gauge theory and gravity amplitudes. \\

\ndt The inverse soft recursion relation for a $n$-point tree-level gauge amplitude is
\bea
\label{IS}
A_n(1,2,.....,n-1,n) = S(n-1,n,1) \,A_{n-1} (1',2.....n-1')\,.
\eea
The prime above indicates that momentum conservation on the right-hand side requires a shift in the momenta of adjacent particles $p_{n-1}$ and $p_1$. For a positive helicity soft particle $n$, the shift is \cite{Boucher-Veronneau:2011rwd}
\bea
\label{s1}
&&|1'\rangle= |1\rangle\,,\, \hspace{1 cm} |1'] = |1] + |n] \frac{\langle n, n-1\rangle}{\langle 1,n-1\rangle} \,,\nn\\
&&|n-1'\rangle= |n-1\rangle\,,\, \hspace{1 cm} |n-1'] = |n-1] + |n] \frac{\langle n 1\rangle}{\langle n-1, 1\rangle} \,.
\eea
\ndt  Here only the neighbouring particles are shifted because only they are  affected by the soft limit. For the case of gravity, the soft factor depends on all legs thus inverse soft expression (\ref{IS}) will involve sum over all particles. \\

\ndt We will now employ the inverse soft method to construct higher-point amplitudes for theories with higher-dimensional operators. We use the three-point amplitudes found in the previous section as seed amplitudes to systematically construct different classes of higher-point amplitudes.

\subsection{$HF^2$ operator}
This is the simplest possible gauge invariant higher-dimensional operator involving spin-0 and spin-1 fields. This is a 5-dimensional operator given by 
\bea
\label{HF}
O^{\alpha'}\,=\,\alpha'\,H\,F_{\mu\nu}^a\,F^{\mu\nu\,a}\,,
\eea

\ndt where $F_{\mu\nu}^{a} = \del_{\mu} A_{\nu}^{a} - \del_{\nu} A^{a}_{\mu} + g\,f^{abc} A^b_{\mu} A^{c}_{\nu}$ is the gluon field strength and $f^{abc}$, the structure constant of the gauge group. $H$ is the real scalar field. We consider the cubic interaction Hamiltonian by plugging $\lambda_1=0\;,\lambda_2=\lambda_3=1$ in (\ref{Hf1})
\bea
\label{Hf}
H^{\alpha'} = \alpha'\int d^3x \; \, \, \phi \,\left[\, \frac{\del^{2}} {\del^{+}} \,\bar{A}^a\, \del^{+}\,\bar{A}^a\,-\,2\,\del\,\bar{A}^a\del\,\bar{A}^a\,+\,\del^{+}\,\bar{A}^a\,\frac{\del^{2}} {\del^{+}} \,\bar{A}^a\,\right] + c.c. \,,
\eea
where $\phi$ is a complex scalar field and $H = \phi + \phi^{\dagger}$. \\

\ndt We can see that by working with the physical fields (using the light-cone gauge), the amplitude for the operator $H F^2$ naturally decomposes into a holomorphic  and an anti-holomorphic parts. This key idea was first presented in \cite{Dixon:2004za} due to the self-duality of the field tensor. The full amplitude for this operator (\ref{HF}) can be obtained as 
\bea
A_n(H,1,2,....n) = A_n (\phi,1,2,....n)+ A_n(\phi^{\dagger},1,2,....n)\,,
\eea
where $A_n$ is the partial color ordered amplitude. The color decomposition for tree-level amplitudes of this operator is similar to the Yang-Mills case \cite{Dixon:2004za}. \\

\ndt   We now construct higher-point amplitudes for this operator. We start with the three-point amplitudes and construct separately the holomorphic and anti-holomorphic amplitudes using the inverse soft and then add them to obtain the full amplitude. \\ 

\ndt The inverse soft recursion relation for this operator is similar to the Yang-Mills case \cite{Boucher-Veronneau:2011rwd} 
\bea
A_{n+1}(\phi,1,2,......n)=S_{YM}\times A_n(\phi,1',2,.....n-1')\,,
\eea
where
\bea
\label{soft}
S_{YM} = \frac{\langle n-1,1 \rangle }{ \langle n-1,n \rangle \langle n 1 \rangle}\,,
\eea
\ndt is the soft factor for the Yang-Mills theory. The non-trivial three-point amplitudes obtained using (\ref{Hf})
\bea
\label{O1}
&&A_3(\phi, 1^- ,2^-) = - \langle 12 \rangle^2\,, \;\;\;\; A_3(\phi^{\dagger}, 1^+ ,2^+) = -[12]^2 \,.
\eea

\ndt  We first construct the MHV amplitudes. The four-point MHV amplitude can be constructed by adding a soft gluon leg of positive helicity to (\ref{O1}). We get
\bea
A_4(\phi, 1^-, 2^-,3^+) &=& - \langle 12 \rangle^2 \times S_{YM} \,, \nn\\
&=& - \langle 12 \rangle^2  \times \frac{\langle 21 \rangle }{ \langle 23 \rangle \langle 3 1 \rangle} \,, \nn\\
&=&  \frac{{\langle 12  \rangle }^3}{ \langle 23 \rangle \langle 3 1 \rangle} \,.
\eea
As $A_4(\phi^{\dagger}, 1^-, 2^-,3^+) =0$, so the full 4-point amplitude is $A_4(H, 1^-, 2^-,3^+) = A_4(\phi, 1^-, 2^-,3^+)$. The above construction is valid only if the amplitude is BCFW constructible. It was shown first in \cite{Berger:2006sh} that the large $z$ behavior of the operator is similar
to the case of Yang-Mills and under a shift of a pair of legs there is no pole at infinity. To see this explicitly, let's consider the following shift  $[1^-, 3^+ \rangle$ and the amplitude $A_4(\phi, 1^-, 2^-,3^+)$ under this shift at large $z$ behaves as
\bea
A_4(\phi, \hat{1}^-, 2^-,\hat{3}^+) =  \frac{{\langle \hat{1}2  \rangle }^3}{ \langle 2\hat{3} \rangle \langle \hat{3} \hat{1} \rangle}=  \frac{{\langle 12  \rangle }^3}{ \langle 2\hat{3} \rangle \langle 3 1 \rangle} \sim \frac{1}{z}\,,
\eea
where we have used $\langle \hat{3} \hat{1} \rangle = \langle {3} {1} \rangle- z \langle {1} {1} \rangle= \langle {3} {1} \rangle$. Therefore the above MHV amplitude for this operator is BCFW constructible and thus inverse soft method is valid. \\

\ndt For 5-point MHV amplitude, we get contribution from both holomorphic and anti-holomorphic part and it can be obtained by multiplying with appropriate soft factor
\bea
A_5(H, 1^-, 2^-,3^+,4^+) &=&  A_5(\phi, 1^-, 2^-,3^+,4^+)+ A_5(\phi^{\dagger}, 1^-, 2^-,3^+,4^+)\,, \nn \\
&=& \frac{{\langle 12  \rangle }^4}{ \langle 12 \rangle \langle 23 \rangle \langle 3 1 \rangle} \times  \frac{\langle 31 \rangle }{ \langle 34 \rangle \langle 4 1 \rangle}\; + \;\frac{[34]^4}{[34] [42][23]} \times \frac{[42]}{[41][12]}\,, \nn \\
&=&  \frac{{\langle 12  \rangle }^4}{ \langle 12 \rangle \langle 23 \rangle \langle 3 4 \rangle \langle 4 1 \rangle} + \frac{[34]^4}{[12] [23][34][41]} \,.
\eea
\ndt This expression matches with the result derived in the literature \cite{Berger:2006sh,Kauffman:1996ix}. \\

\ndt Now beyond 5-point, the calculation for the full MHV amplitude gets tedious as the anti-holomorphic part will have contributions from NMHV side as well. The holomorphic (anti-holomorphic) part is simple and can be constructed recursively by adding a soft gluon leg of positive helicity to a lower-point amplitude.
\bea
A_n(\phi, 1^-, 2^-,3^+,4^+,......,n^+) =  \frac{\langle 12 \rangle^4}{ \langle 12 \rangle \langle 23 \rangle  \langle 34 \rangle.........\langle n1 \rangle} \,.
\eea
\ndt The above formula is valid for both massive and massless scalars, and it reduces to the pure Yang-Mills amplitude when we take the momentum of the scalar to zero. \\

\ndt We now construct the non-MHV amplitudes using the inverse soft method. The four point anti-holomorphic amplitude is constructed by adding a soft gluon of positive helicity and using the appropriate shift defined in (\ref{s1})
\bea
 A_4(\phi^{\dagger}, 1^+ ,2^+,3^+) &=& -[1'2']^2 \times \frac{\langle 21 \rangle }{ \langle 23 \rangle \langle 3 1 \rangle} \,,\nn\\
 &=& \left( [12] + [13]  \frac{\langle 31 \rangle}{\langle 21 \rangle} + [32] \frac{\langle 32 \rangle}{\langle 12 \rangle} \right)^2 \times \frac{\langle 21 \rangle }{ \langle 23 \rangle \langle 3 1 \rangle}\,,\nn\\ \nn \\
&=& \frac{\Big(\langle 12 \rangle [21] + \langle 23 \rangle [32] +  \langle 13 \rangle [31]\Big)^2}{\langle 12 \rangle\langle 23 \rangle\langle 31 \rangle}\,.
\eea
One can now recursively construct this for $n$-point by multiplying with appropriate soft factor and deforming the adjacent legs
\bea
A_n(\phi^{\dagger}, 1^+ ,2^+,......,n^+) &=& \frac{\left( \sum_{1 \leq i <j \leq n-1} \langle ij \rangle [j'i'] \right)^2 }{ \langle 12 \rangle\langle 23 \rangle ..... \langle n-1,1 \rangle} \times  \frac{\langle n-1,1 \rangle }{ \langle n-1,n \rangle \langle n 1 \rangle}\,, \nn\\
&=& \frac{\left( \sum_{1 \leq i <j \leq n} \langle ij \rangle [ji] \right)^2 }{ \langle 12 \rangle\langle 23 \rangle ....... \langle n-1,n \rangle\langle n1 \rangle}\,.
\eea
\ndt The above amplitude under the shift $[1^+, n^+ \rangle$ behaves well at large $z$ and there is no pole at infinity so the inverse soft construction is valid for non-MHV amplitudes. As, $A_3(\phi, 1^+ ,2^+)=0$ therefore all higher-point amplitudes will be zero. So the full $n$-point amplitude is 
\bea
 A_n(H, 1^+,2^+,....n^+) = \frac{\left( \sum_{1 \leq i <j \leq n} \langle ij \rangle [ji] \right)^2 }{ \langle 12 \rangle\langle 23 \rangle ....... \langle n-1,n \rangle\langle n1 \rangle}\,.
\eea
Similarly $ A_n(H, 1^-,2^-,....n^-)$ can be constructed using the inverse soft method. Our results match with the amplitudes derived using MHV vertex expansion method in the literature \cite{Dixon:2004za,Berger:2006sh}.  \\
 
\subsection{$F^3$ operator}
We now consider a gauge invariant higher-dimensional operator involving purely spin-1 fields. The simplest of such operators is 6-dimensional given by
\bea
\label{O2}
O^{\alpha^2} = \alpha^2\,f^{abc} F_{\mu}^{a \;\nu}\, F_{\nu}^{b \;\rho}\,F_{\rho}^{c \;\mu}\ ,
\eea
where $F_{\mu}^{a \;\nu} = \del_{\mu} A^{a\,\nu} - \del^{\nu} A^{a}_{\mu} + g\,f^{abc} A^b_{\mu} A^{c\,\nu}$ is the gluon field strength and $f^{abc}$, the structure constant of the gauge group. 
\vskip 0.3cm
\ndt The cubic interaction vertex can be obtained from (\ref{Hf1}) by setting $\lambda_1=\lambda_2=\lambda_3=1$ so $\lambda=3$, and the Hamiltonian is
\bea
\label{F3}
H^{\alpha^2} &=& \alpha^2\int d^3x\, f^{abc} \,\bigg\{ \frac{1}{\del^+} \bar{A}^ a \frac{\del^3}{\del^{+}} \bar{A}^b \del^{+\,2} \bar{A}^c - \frac{1}{\del^+} \bar{A}^ a \del^{+\,2} \bar{A}^b \frac{\del^3}{\del^{+}} \bar{A}^c - 3 \frac{1}{\del^+} \bar{A}^ a {\del^2} \bar{A}^b \del \del^{+} \bar{A}^c \nn\\
 &+& 3\, \frac{1}{\del^+} \bar{A}^ a \del \del^{+} \bar{A}^b {\del^2} \bar{A}^c \bigg\} + c.c. \,.
\eea

\ndt Similar to the previous case, we see that the decomposition of the operator (\ref{O2}) into holomorphic and anti-holomorphic parts is manifest in the light-cone gauge. This decomposition was first presented in \cite{Dixon:2004za,Broedel:2012rc}. So the full $F^3$ amplitude can be constructed by taking the sum of the holomorphic and anti-holomorphic amplitude. \\

\ndt In this theory, holomorphic amplitudes with exactly three negative helicity gluons and an arbitrary number of positive helicity gluons are referred as MHV amplitudes and are denoted by $A^{F_+}$. The anti-holomorphic amplitudes with exactly three positive helicity gluons and an arbitrary number of negative helicity gluons are referred as anti-MHV and are denoted by $A^{F_-}$.  \\

\ndt  It was shown in \cite{Elvang:2016qvq,Bianchi:2014gla} that for the higher derivative theory of massless particles in four dimensions, the tree-level soft photon and graviton theorems receive modifications at subleading and subsubleading orders. However, the leading soft factors are not altered for these theories because these interactions ($F^3,R^3$) are generically suppressed in the soft-limit~\cite{Bianchi:2014gla}. Therefore, the leading soft factor for the Yang-Mills theory with $F^3$ correction in the soft gluon limit is same as (\ref{soft}).\\

\ndt The three-point amplitudes extracted from (\ref{F3}) are
\bea
 &&A^{F^3}_3(p^-,\,k^-,\,l^-)\,\equiv\, A^{F_+}_3(p^-,\,k^-,\,l^-)=\langle k l \rangle \langle l p \rangle\langle pk  \rangle\,, \nn\\ \nn\\
&&A^{F^3}_3(p^+,\,k^+,\,l^+) \,\equiv\, A^{F_-}_3(p^+,\,k^+,\,l^+)= [k l] [ l p ][ pk ]\,.
\eea
\ndt A four-point tree-level MHV amplitude corresponding to a single insertion of the $F^3$ operator can be constructed using this as follows (we attach a soft gluon between the adjacent legs $1$ and $3$).
\bea
A^{F^3}_4(1^-, 2^-,3^-,4^+) &=& A^{F_+}_3(1^-, 2^-,3^-) \times S_{YM} \,,\nn\\
&=& \langle 12 \rangle \langle 23 \rangle \langle 31 \rangle \times  \frac{\langle 31 \rangle }{ \langle 3 4 \rangle \langle 4 1 \rangle} \,, \nn\\
&=& \frac{\langle 12 \rangle^2 \langle 23 \rangle^2 \langle 31 \rangle^2}{ \langle 12 \rangle \langle 23 \rangle  \langle 34 \rangle  \langle 41 \rangle} \,.
\eea
\ndt The above construction is valid because under the following shift $[1^-, 4^+ \rangle$, the amplitude has no pole at infinity. So the inverse soft construction is valid for this class of amplitudes. The 5-point MHV amplitude can also be constructed similarly. At the 6-point level, the full $F^3$ amplitude will get contributions from both holomorphic and anti-holomorphic parts
\bea
A^{F^3}_6(1^-, 2^-,3^-,4^+,5^+,6^+) &=& A^{F_+}_6(1^-, 2^-,3^-,4^+,5^+,6^+) + A^{F_-}_6(1^-, 2^-,3^-,4^+,5^+,6^+)\,, \nn\\ \nn\\
&=&  \frac{\langle 12 \rangle^2 \langle 23 \rangle^2 \langle 31 \rangle^2}{ \langle 12 \rangle \langle 23 \rangle  \langle 34 \rangle  \langle 45 \rangle \langle 56\rangle \langle 61 \rangle} + \frac{[45]^2[56]^2[64]^2}{ [12][23][34][45][56][61]}\,.
\eea
Beyond this order, the calculation for the complete $F^3$ amplitude becomes complicated since not {\it {all}} relevant seed amplitudes are constructible within this formalism. \\

\ndt To any order, the MHV holomorphic amplitudes may be constructed recursively - within this formalism - by attaching soft gluon legs to lower point amplitudes.
\bea
\label{An}
A^{F_+}_n(1^-, 2^-,3^-,4^+,......,n^+) =  \frac{\langle 12 \rangle^2 \langle 23 \rangle^2 \langle 31 \rangle^2}{ \langle 12 \rangle \langle 23 \rangle  \langle 34 \rangle.........\langle n1 \rangle} \,.
\eea 
\ndt Note that for $n \leq 6$, this is *also* the full MHV amplitude. Similarly $A^{F_-}_n(1^+, 2^+,3^+,4^-,......,n^-)$ can be obtained from (\ref{An}) by parity. The above expression matches with the results in \cite{Dixon:1993xd, Dixon:2004za} obtained using Feynman rules.  \\

\ndt Another class of amplitudes that may be built in this formalism are the NMHV amplitudes. For these, we start with the seed amplitude $A_4^{F_+}(1^-,2^-,3^-,4^-)$~\cite{Broedel:2012rc,Cohen:2010mi}
\bea
A_{4}^{F_+}(1^-,2^-,3^-,4^-)  = 2 \frac{stu}{[12][23][34][41]}\ ,
\eea
\ndt where $s,t,u$ are the usual Mandelstam variables. For example, we can compute the $5$-point amplitude
\bea
A_{5}^{F_+}(1^-,2^-,3^-,4^-,5^+)&=& S(5,1,2) A_4(2'^-,3^-,4^-,5'^+) + S(3,4,5) A_4(1^-,2^-,3'^-,5'^+) \nn
\\ 
&&\hspace{1 cm} + S(4,5,1) A_4(1'^-,2^-,3^-,4'^-) \,, \nn\\ \nn\\
&=& 	\frac{\langle 34 \rangle^2 [35]^2 [45]}{[12][23][34][15]} + \frac{\langle 12 \rangle^2 [25]^2 [15]}{[12][23][34][45]} + 2\frac{\langle14 \rangle^2 \langle 13 \rangle \langle 24 \rangle}{[23] \langle 45 \rangle \langle 15 \rangle }\,.
\eea 
\ndt This matches with the result in \cite{Broedel:2012rc}. As pointed out in \cite{Boucher-Veronneau:2011rwd} this technique will only work up to $n \leq 8$ for gauge theory and for some classes of NMHV amplitudes with arbitrary number of legs. 

\subsection{$R^3$ operator}
We now consider a higher-dimensional operator in the case of gravity denoted by $R^3$ . This operator also decomposes into holomorphic and anti-holomorphic part as explained before. In this section, we focus only on the holomorphic amplitude denoted by $M^{R_+}$. The construction for anti-holomorphic part follows similarly. The full amplitude is again a sum of contributions from both holomophic and anti-holomorphic parts. We start with (\ref{V1}) and set $\lambda_1=\lambda_2=\lambda_3=2$, so the three-point holomorphic amplitude is
 \bea
M^{R_+}_3(p^-,\,k^-,\,l^-) = \langle 12 \rangle^2 \langle 23 \rangle^2 \langle 31  \rangle^2 \, .
\eea 
\ndt The inverse soft recursion relation for gravity, for a $n$-point amplitude, reads \cite{Boucher-Veronneau:2011rwd}
\bea
M_{MHV}(1,.......,n^+) = \sum_{i=3}^{n-1} \frac{[in]\langle 1i \rangle \langle 2i \rangle}{\langle ni \rangle \langle 1n \rangle \langle 2n \rangle} M_{MHV} (1',....,i',...., (n-1))\,.
\eea
The prime indicates the momentum shift, mentioned earlier. For a positive helicity soft graviton $k$, the shift is
\bea
\label{Gs}
&&|i'\rangle= |i\rangle\,,\, \hspace{1 cm} |i'] = |i] + |k] \frac{\langle k j\rangle}{\langle i j\rangle} \,,\nn\\
&&|j'\rangle= |j\rangle\,,\, \hspace{1 cm} |j'] = |j] + |k] \frac{\langle k i\rangle}{\langle  ji\rangle} \,.
\eea Using the inverse soft method, the four-point tree-level amplitude is
\bea
\label{M4}
M^{R_+}_4(1^-, 2^-,3^-,4^+) &=&  \langle 12 \rangle^2 \langle 23 \rangle^2 \langle 31 \rangle^2 \times \frac{[34]\langle 13 \rangle \langle 23 \rangle}{\langle 43 \rangle \langle 14 \rangle \langle 24 \rangle}\,,\nn\\
&=& \frac{ \langle 12 \rangle^3 \langle 23 \rangle^3 \langle 31 \rangle^3\, [12] }{\langle 34 \rangle^2 \langle 14 \rangle \langle 24 \rangle} \,.
\eea
\ndt For the case of gravity and $R^3$ theories, the higher-point amplitudes are functions of both the holomorphic and anti-holomorphic spinors that makes the construction of amplitudes using the inverse soft approach more involved. We show below an example of the construction of the five-point amplitude in the $R^3$ theory.
\bea
M^{R_+}_5(1^-, 2^-,3^-,4^+,5^+) &=& \sum_{i=3}^{4} \frac{[i5]\langle 1i \rangle \langle 2i \rangle}{\langle 5i \rangle \langle 15 \rangle \langle 25 \rangle} M^{R_+}_4 (1'^-,2^-,..., i') \,,\nn\\
&=& \frac{[35]\langle 13 \rangle \langle 23 \rangle}{\langle 53 \rangle \langle 15 \rangle \langle 25 \rangle} M^{R_+}_4 (1'^-,2^-,3'^-,4^+) + \frac{[45]\langle 14 \rangle \langle 24 \rangle}{\langle 54 \rangle \langle 15 \rangle \langle 25 \rangle} M^{R_+}_4 (1'^-,2^-,3^-,4'^+) \,. \nn
\eea
Using the appropriate shifts, the amplitude reads
\bea
M^{R_+}_5(1^-, 2^-,3^-,4^+,5^+)&=&\frac{ \langle 12 \rangle^4 \langle 23 \rangle^4 \langle 31 \rangle^4}{\langle 12 \rangle \langle 13 \rangle \langle 14 \rangle\langle 34 \rangle \langle 15 \rangle \langle 25 \rangle} \left[ \frac{[35][24]}{\langle 35 \rangle \langle 24\rangle} - \frac{[45][23]}{\langle 45 \rangle \langle 23 \rangle} \right]\,, \nn\\
&=& \frac{ \langle 12 \rangle^4 \langle 23 \rangle^4 \langle 31 \rangle^4}{\langle 12 \rangle^8}\,M^{R}_5(1^-, 2^-,3^+,4^+,5^+)\,,
\eea
\ndt where $M_5^{R}$ is the five-point amplitude for pure gravity (the proportionality to gravity ensures `constructibility'). The $n$-point holomorphic MHV amplitude for $R^3$ operator can be written as
\bea
M^{R_+}_n(1^-, 2^-,3^-,4^+,.......,n^+)=\frac{ \langle 12 \rangle^4 \langle 23 \rangle^4 \langle 31 \rangle^4}{\langle 12 \rangle^8}\,M^{R}_n(1^-, 2^-,3^+,.......,n^+) \,.
\eea
\ndt The anti-holomorphic part corresponding to $R^3$ operator can be similarly obtained. This is consistent with the results derived in \cite{Broedel:2012rc} using color-kinematic duality. Similar to the previous section, we can construct a class of higher-point NMHV amplitudes using the inverse soft technique. We present the construction of the five-point NMHV amplitude in appendix D.

\section*{Acknowledgments}
We thank Mohd Ali for his useful insights and discussion. The work of SA is partially supported by a MATRICS grant - MTR/2020/000073 - of SERB. SP and NB acknowledge support from the Prime Minister’s Research Fellowship (PMRF) and CSIR-NET fellowship respectively. \\

\appendix
\section*{Appendix}
\vskip 0.2 cm
\section{Light-cone Poincar\'e invariance}

\ndt The generators of the Poincar\'e  algebra, in the light-cone coordinates, are the momenta
\bea
	p^+=\partial^{+}=-p_-\,, \hspace{ 0.4 cm} p^-=\frac{\partial \bar{\del}}{\del^+}=-p_+\,, \hspace{ 0.4 cm} p=\partial, \hspace{ 0.4 cm}\bar{p}=\bar{\partial} \,,
\eea
the rotation generators 
\bea
        && j= (x\bar{\partial}-\bar{x}\partial - \lambda)\,, \hspace{ 1.3 cm}     j^{+-}=\left(x^+\frac{\partial \bar{\del}}{\del^+}-x^-\partial^{+}   \right)\,, \nn \\
	&& j^+=\left(x^+\partial-x\,{\partial}^+ \right),  \hspace{ 1 cm}  j^-=\left(x^-\partial-x\,\frac{\partial \bar{\del}}{\del^+} +\lambda \frac{\del}{\del^+} \right) \,, \nn
\eea
and their complex conjugate. $\lambda$ is the helicity of the field and $\partial^{-}=\frac{\partial\bar{\partial}}{\partial^{+}}$ for a free theory (modified by corrections when interactions are switched on).\\

\ndt The Hamiltonian for the free field theory is
\begin{equation}
H\equiv\int d^{3}x\,\mathcal{H}=-\int d^3x\,\bar\phi\,{\partial\bar\partial}\,\phi\  ,
\end{equation}
with the second equality being valid only for the free theory. We also write
\begin{equation}
\label{hamil}
H\equiv\int d^{3}x\,\mathcal{H}=\int d^{3}x\,\partial_{-}\bar{\phi}\,\delta_{p^-}\phi\ ,
\end{equation}
in terms of the time translation operator
\begin{eqnarray}
\delta_{p^-}\phi \equiv\partial_{+}\phi=\lbrace \phi,\mathcal{H}\rbrace\ .
\end{eqnarray} 

\ndt When interactions are switched on, this $\delta_{p^-} \phi$ operator picks up corrections, order by order in the coupling constant $g$. The detailed derivation of light-cone cubic interaction vertices for arbitrary spin theories are presented in~\cite{BBB1, Akshay:2014qea}. Here, we give schematic proof to derive the interaction vertices by the closure of Poincar\'e algebra.  We start with the ansatz for the cubic interaction vertices based on helicity considerations and dimensional analysis. The ansatz reads
\bea
\label{h0}
\delta^{g}_{{p}^{-}} \phi_1  =g\, A \,\del^{+\,\mu} \left[\, \bar{\del}^a\,\del^{+\, \rho} {\phi}_2\, \bar{\del}^b\, \del^{+\, \sigma} {\phi}_3\,\right] \,,  
\eea
\ndt  where the fields ${\phi}_1,{\phi}_2$ and ${\phi}_3$ have integer spins $\lambda_1, \lambda_2$ and $\lambda_3$, the dimension of the coupling constant is $[g] =L^{\lambda_2 +  \lambda_3  -  \lambda_1- 1}$ and  $\mu, \rho,\sigma, a,b$ are integers. \\

\ndt As we turn on interactions, the other dynamical generators also pick up corrections
\bea
&&\delta_{j^{+-}} \phi = \delta^0_{j^{+-}} \phi + x^+ \delta^{g}_{{p}^{-}} \phi + O(g^2) \,, \nn\\
&& \delta_{j^{-}} \phi = \delta^0_{j^{-}} \phi -x \delta^{g}_{{p}^{-}} \phi + \delta^g_s \phi + O(g^2) \,,\nn\\
&& \delta_{\bar{j}^{-}} \phi = \delta^0_{\bar{j}^{-}} \phi -\bar{x} \delta^{g}_{{p}^{-}} \phi + \delta^g_{\bar{s}} \phi +  O(g^2) \,,
\eea
where $\delta^g_s$ and $\delta^g_{\bar{s}}$ represent the spin transformations.\\

\ndt The requirement of closure of the Poincar\'e algebra imposes various conditions on the integers introduced in (\ref{h0}). The result is 
\begin{eqnarray}
\delta_{p^-}^{g}\phi_1 = g \sum^{\lambda}_{n=0} (-1)^{n}{\lambda \choose n}\partial^{+\,(\lambda_1 -1)}\left[{\bar{\partial}^{n}}\,{\partial^{+\,(\lambda-\lambda_2-n)}}\,\phi_2\, {\bar{\partial}^{(\lambda- n)}}{\partial^{+\,(n-\lambda_3)}}\phi_3\right] \,.
\end{eqnarray}
 
\ndt From (\ref {hamil}), the complete Hamiltonian to this order reads~\cite{Akshay:2014qea}
\bea
\label{H00}
H^{g} = g\int d^3x \; \,\sum_{n=0}^{\lambda}\, (-1)^n {\lambda \choose n}\,  \bar{\phi}_1 \,{\del^{+\,\lambda_1}}\,\left[{\bar{\partial}^{n}}\,{\partial^{+\,(\lambda-\lambda_2-n)}}\,\phi_2\, {\bar{\partial}^{(\lambda- n)}}{\partial^{+\,(n-\lambda_3)}}\phi_3\right] + c.c. \,.
\eea
\\
\ndt If we set $\lambda_1 = \lambda_2 = \lambda_3 = \lambda'$ in (\ref{H00}) with $\lambda'$ odd, $H^{g}$ vanishes. Hence a self-interaction Hamiltonian for odd integer spins exists, if and only if we introduce a gauge group and it reads
\bea
\label{odd}
H=\int d^{3}x  \left(\partial \bar{\phi}^{a}\bar{\partial}\phi^a-g f^{abc}\sum^{\lambda'}_{n=0} (-1)^{n}{\lambda' \choose n}\bar\phi^a\;\partial^{+\,\lambda'}\,\left[\frac{\bar{\partial}^{(\lambda' -n)}}{\partial^{+\,(\lambda' -n)}}\phi^b\, \frac{\bar{\partial}^{n}}{\partial^{+\,n}}\phi^{c}\right]+\mathcal{O}(g^2)\right) .
\eea
\\

\section{Derivation of cubic vertices with mixed derivatives}
\ndt The refined ansatz for cubic interaction vertices with mixed derivatives reads
 \bea
 \label{t2v}
\delta^{''\,\alpha}_{{p}^{-}} \phi_1  = \alpha\,\sum_{n=0}^{\lambda_2 +  \lambda_3 }\,\sum_{m=0}^{\lambda_1}\, C_{n,m} \,\del^{+\,\mu_{n,m}} \left[\, \bar{\del}^{(\lambda_2 +  \lambda_3 -n)}\,\del^{(\lambda_1-m)} \,\del^{+\, \rho_{n,m}}\, {\phi}_2\, \bar{\del}^n\,\del^m \,\del^{+\, \sigma_{n,m}}\, {\phi}_3\,\right]  .
\eea
\ndt To determine the vertex, we first use the kinematical commutators 
\bea
[\,\delta_{{j}^+} \,, \delta^{''\alpha}_{{p}^{-}}\,]\phi_1 = 0\,, \;\;\;\;\ [\,\delta_{\bar{j}^{+}} \,, \delta^{''\alpha}_{{p}^{-}}\,]\phi_1 = 0\,.
\eea
They give the following conditions 
\bea
&&\sum_{n=0}^{\lambda_2+\lambda_3 }\,\sum_{m=0}^{\lambda_1}\; C_{n,m} \bigg\{  \,(\lambda_2+\lambda_3  -n)\,\del^{+\,\mu_{n,m}} \,\left[\, \bar{\del}^{(\lambda_2+\lambda_3 -n-1)}\,\del^{(\lambda_1-m)} \,\del^{+\, (\rho_{n,m} + 1)} \,{\phi}_2\, \bar{\del}^n\,\del^m \del^{+\, \sigma_{n,m}} \,{\phi}_3\,\right]  \nn\\
&& \hspace{1.8 cm}+ \, n \,\del^{+\,\mu_{n,m}} \,\left[\, \bar{\del}^{(\lambda_2+ \lambda_3 -n)}\,\del^{(\lambda_1-m)}\, \del^{+\, \rho_{n,m} }\, {\phi}_2\, \bar{\del}^{(n-1)}\,\del^m \del^{+\, (\sigma_{n,m}+1)} \,{\phi}_3\,\right] \bigg\}=0 \,,
\eea
\bea
&&\sum_{n=0}^{\lambda_2+\lambda_3 }\,\sum_{m=0}^{\lambda_1}\;C_{n,m} \bigg\{  \,(\lambda_1  -m)\,\del^{+\,\mu_{n,m}} \,\left[\, \bar{\del}^{(\lambda_2+\lambda_3 -n)}\,\del^{(\lambda_1-m-1)} \,\del^{+\, (\rho_{n,m} + 1)} \,{\phi}_2\, \bar{\del}^n\,\del^m \del^{+\, \sigma_{n,m}}\, {\phi}_3\,\right]  \nn\\
&& \hspace{1.2 cm}+ \, m \,\del^{+\,\mu_{n,m}} \,\left[\, \bar{\del}^{(\lambda_2+\lambda_3  -n)}\,\del^{(\lambda_1-m)}\, \del^{+\, \rho_{n,m} }\, {\phi}_2\, \bar{\del}^{n}\,\del^{(m-1)} \del^{+\, (\sigma_{n,m}+1)} \,{\phi}_3\,\right] \bigg\}=0\,.
\eea
\ndt These conditions are satisfied if the coefficients obey the following recursion relations
\bea
\label{c1}
&&C_{n+1,m} = - \frac{(\lambda_2+ \lambda_3 - n)}{(n+1)}\, C_{n,m}\;=\; (-1)^{n+1} {\lambda_2+\lambda_3 \choose n+1}\,C_{0,m} \;, \nn\\
&&C_{n,m+1} = - \frac{(\lambda_1 - m)}{(m+1)}\, C_{n,m} \;=\; (-1)^{m+1} {\lambda_1 \choose m+1}\,C_{n,0} \;, \\ \nn\\
\label{c3}
&& \rho_{n+1,m} = \rho_{n,m}+1 \,,\;\;\; \sigma_{n+1,m} = \sigma_{n,m}-1 \,,\;\;\; \mu_{n+1,m} = \mu_{n,m}\;,\nn\\
&& \rho_{n,m+1} = \rho_{n,m}+1 \,,\;\;\; \sigma_{n,m+1} = \sigma_{n,m}-1 \,,\;\;\; \mu_{n,m+1} = \mu_{n,m}\;.
\eea

\vskip 2 mm
\ndt The idea is to use the dynamical commutators
 $[\,\delta_{{j}^{-}} \,, \delta^{''}_{{p}^{-}}\,]^{\alpha}\,\phi_1 = 0$,\;$[\,\delta_{\bar{j}^{-}} \,, \delta^{''}_{{p}^{-}}\,]^{\alpha}\,\phi_1 = 0$ to fix $\rho,\mu,$ and $\sigma$. To solve these commutators we need the spin parts  $\delta_{{s}}^{\alpha}\phi\,, \delta_{\bar{s}}^{\alpha}\phi$. For type-2 vertices, these are structurally 
\bea
\delta_{{s}}^{\alpha}\phi_1 &\sim& \bar{\del}^{\lambda_2+\lambda_3 -1 }\del^{\lambda_1}\phi_2 \, \phi_3 + \bar{\del}^{\lambda_2-1} \del^{\lambda_1+\lambda_3} \phi_2 \,\bar{\phi}_3\,, \\
 \delta_{\bar{s}}^{\alpha}\phi_1 &\sim& \bar{\del}^{\lambda_2+\lambda_3 } \del^{\lambda_1-1} \phi_2  \, \phi_3 + \bar{\del}^{\lambda_2} \del^{\lambda_1+\lambda_3-1} \phi_2 \, \bar{\phi}_3\,.
\eea
\vskip 2 mm
\ndt For type-2 vertices, both the spin parts have a non-trivial structure and closing the dynamical commutators  $[\,\delta_{{j}^{-}} \,, \delta^{''}_{{p}^{-}}\,]^{\alpha}\,\phi_1 = 0$\;\,,\;\,$[\,\delta_{\bar{j}^{-}} \,,\,\delta^{''}_{{p}^{-}}\,]^{\alpha}\,\phi_1 = 0$ does not fix the values of $\rho,\mu,$ and $\sigma$. Therefore we are forced to introduce two functions $u (\lambda_i)$ and $v(\lambda_i)$ that capture this ambiguity in the powers of $\partial^+$.\\

\ndt Plugging the values of the constants from equation (\ref{c1}),(\ref{c3}) in our refined ansatz (\ref{t2v}), we find
\bea
\delta^{''\,\alpha}_{{p}^{-}} \phi_1 & =& \alpha\,\sum_{n=0}^{\lambda_2 +  \lambda_3 }\,\sum_{m=0}^{\lambda_1}\, (-1)^{(n+m)}\,{\lambda_2+\lambda_3 \choose n}\,{\lambda_1 \choose m} \nn\\
&& \bigg\{\,\del^{+\,{\mu} }  \left[\, \bar{\del}^{(\lambda_2 +  \lambda_3 -n)}\,\del^{(\,\lambda_1-m)} \,\del^{+\, {\left(\,n+m +u \right)}} \,{\phi}_2\, \bar{\del}^n\,\del^m \,\del^{+\, {\left(v- n-m\,\right)}} \,{\phi}_3\,\right] \bigg\} \,, \nn
\eea
with $ n+m + u= \rho_{n,m}\,\,,\,\, v-n-m=\sigma_{n,m}$ and $\mu=\mu_{n,m}$.\\

\ndt Since
\bea
H = \int d^3x \; \del^+ \bar{\phi}_1 \,\delta^{''\alpha}_{{p}^{-}} \phi_1 \,, \nn
\eea
the interaction Hamiltonian is 
\bea
\label{H222}
H^{\alpha} &=& \alpha\int d^3x \; \sum_{n=0}^{\lambda_2 +  \lambda_3 }\,\sum_{m=0}^{\lambda_1}\, (-1)^{(n+m)}\,{\lambda_2+\lambda_3 \choose n}\,{\lambda_1 \choose m} \; \nn\\ 
&& \!\!\!\!\bigg\{  {\del^{+\,(\mu +1)}} \,\bar{\phi}_1\,\bar{\del}^{(\lambda_2 +  \lambda_3 -n)}\,\del^{(\,\lambda_1-m)} \,\del^{+\, {\left(\,n+m+u\right)}} \,{\phi}_2\, \bar{\del}^n\,\del^m \,\del^{+\, {\left(v- n-m\,\right)}} \,{\phi}_3\, \bigg\} + c.c. \;.  \nn \\
\eea

\section{Quartic vertices in higher derivative theories}

\ndt In this appendix, we demonstrate the construction of quartic interaction vertices using closure of Poincar\'e algebra.  We consider the case where $\lambda_1=0,\lambda_2=1,\lambda_3=1,\lambda_4=1$. The possible helicity configurations for this vertex are: $ (0---)\,,\,(0+++)\,,\, (0++-)\,,\, (0--+)$. We first construct the vertex for the helicity configuration $(0---)$.  \\

\ndt The ansatz for the quaric vertex is

\bea
\delta_{p^-}^{\beta}\phi=\beta\,X\; f^{ijk}\, \partial^{\,a}\partial^{+\,\mu}\bigg\{\partial^{\,b}\partial^{+\,\rho}\bar{A}^i\,\partial^{\,c}\partial^{+\,\sigma}\bar{A}^j\,\partial^{\,d}\partial^{+\,\delta}\bar{A}^k\,\bigg\}\,,
\eea

\ndt where $\beta$ is the coupling constant, $X$ is a constant and $f^{ijk}$ is the structure constant. The commutator $[\,j\,,\,p^-\,]^{\beta}$ produces
\bea
\label{partition}
a+b+c+d=3\,.
\eea

\ndt The commutator $[\,\bar{j}^{\,+}\,,\,p^-\,]^{\beta}=0$ yields

\bea
&&X(a,b,c)\bigg\{a\,\partial^{\,(a-1)}\partial^{+\,(\mu+1)}\left[\partial^{\,b}\partial^{+\,\rho}\bar{A}\,\partial^{\,c}\partial^{+\,\sigma}\bar{A}\,\partial^{\,d}\partial^{+\,\delta}\bar{A}\right]+\nn\\
&&b\;\partial^{\,a}\partial^{+\,\mu}\left[\partial^{\,(b-1)}\partial^{+\,(\rho+1)}\bar{A}\,\partial^{\,c}\partial^{+\,\sigma}\bar{A}\,\partial^{\,d}\partial^{+\,\delta}\bar{A}\right]+c\,\partial^{\,a}\partial^{+\,\mu}\left[\partial^{\,b}\partial^{+\,\rho}\bar{A}\,\partial^{\,(c-1)}\partial^{+\,(\sigma+1)}\bar{A}\,\partial^{\,d}\partial^{+\,\delta}\bar{A}\right]\nn\\
&&+\,d\;\partial^{\,a}\partial^{+\,\mu}\left[\partial^{\,b}\partial^{+\,\rho}\bar{A}\,\partial^{\,c}\partial^{+\,\sigma}\bar{A}\,\partial^{\,(d-1)}\partial^{+\,(\delta+1)}\bar{A}\right]\bigg\}=0.\nn\\
\eea

\ndt Let $a\equiv a+1$ in term $1$, $b\equiv b+1$ in term $2$, $c\equiv c+1$ in term $3$ and $d=3-a-b-c$ in term $4$. This generates the following set of recursion relations

\bea
\label{recur}
&&\bigg\{(a+1)X(a+1,b,c)+(b+1)X(a,b+1,c)+(c+1)X(a,b,c+1)\nn\\
&&+(3-a-b-c)X(a,b,c)\bigg\}=0\nn\\
&&\mu(a+1)=\mu(a)-1\,\,\, \,;\,\,\, \, \rho(a+1)=\rho(a)-1\,\,\,\,;\,\,\,\,\sigma(c+1)=\sigma(c)-1\nn\\
&&\delta(a+1,b,c)=\delta(a,b,c)+1\;\;;\;\;\delta(a,b+1,c)=\delta(a,b,c)+1\nn\\
&&\delta(a,b,c+1)=\delta(a,b,c)+1\,.\nn\\
\eea
\ndt The solutions to the recursion relation fall in four independent classes.  For this helicity configuration, $\delta_s^{\beta}\phi$ does not exist because its consistency with the helicity generator $j$ requires such a term to have four transverse derivatives rendering it inconsistent with the dimensionality. We then fix the values of $\mu\,,\,\rho\,,\,\sigma\,,\,\delta$ using the commutator with dynamical generator $j^-$ for each class of solution. The final form of the quartic vertex is
\bea
\label{pF}
H^{\beta}&=&\beta\;\,f^{ijk}\;\Bigg\{\,\int d^3x \; \,\sum_{N=0}^{3}\,\sum_{b=0}^{N}\,\sum_{c=0}^{3-N}\,(-1)^N\,(-1)^{b+c}\,{N\choose b}\,{3-N\choose c}\nn\\ \nn\\
&&\, C_1\; \phi\;\partial^{\,b}\partial^{+\,(N-1-b)}\bar{A}^i\,\partial^{\,c}\partial^{+\,(2-N-c)}\bar{A}^j\,\partial^{\,(3-b-c)}\partial^{+\,(b+c-1)}\bar{A}^k\nn\\ 
&+&\int d^3x \; \,\sum_{N=0}^{3}\,\sum_{a=0}^{N}\,\sum_{c=0}^{3-N}\,(-1)^N\,(-1)^{a+c}\,{N\choose a}\,{3-N\choose c}\, \nn\\
&& \bigg\{\,C_2\;\partial^{\,a}\partial^{+\,(N-a)}\phi\;\frac{1}{\partial^+}\bar{A}^i\;\partial^{\,c}\partial^{+\,(2-N-c)}\bar{A}^j\,\partial^{\,(3-a-c)}\partial^{+\,(a+c-1)}\bar{A}^k\nn\\ \nn\\
&+&\,C_3\;\partial^{\,a}\partial^{+\,(N-a)}\phi\;\partial^{\,(N-a)}\partial^{+\,(a-1)}\bar{A}^i\, \partial^{\,c}\partial^{+\,(2-N-c)}\bar{A}^j\,\partial^{\,(3-N-c)}\partial^{+\,(c-1)}\bar{A}^k\nn\\
&+&\,C_4\,\partial^{\,(3-a-c)}\partial^{+\,(a+c-1)}\phi\;\frac{1}{\partial^{+}}\bar{A}^i\; \partial^{\,a}\partial^{+\,(2-N-a)}\bar{A}^j\,\partial^{\,c}\partial^{+\,(N-1-c)}\bar{A}^k\bigg\}\,\Bigg\} \,+ c.c.
\eea
where $C_1,C_2,C_3$ and $C_4$ are numerical cofficients to be fixed. The Poincar\'e algebra uniquely fixes the structure of the vertex up to the overall numerical constant for each possible solution. These coefficients will be fixed using the fact that the vertex is antisymmetric under the exchange of gluon legs. \\

\ndt The form of the vertex (\ref{pF}) in real space is complicated, we write the vertex in the momentum space such that it is manifestly antisymmetric.  In momentum space, the quartic vertex has the following structure (measure and constants suppressed)
\bea
H^{\beta} &=& f^{ijk} \sum_{N=0}^{3}\,\sum_{a=0}^{N}\,\sum_{c=0}^{3-N}\,(-1)^N\,(-1)^{a+c}\,{N\choose a}\,{3-N\choose c}\nn\\ \nn\\
&&\bigg\{\,C_1 \;p^{\,a}\,p^{+\,(N-1-a)}\,k^{\,c}\,k^{+\,(2-N-c)}\,l^{\,(3-a-c)}\, l^{+\,(a+c-1)}\nn\\ 
&+&\, C_2\; q^{\,a}\,q^{+\,(N-a)}\;\frac{1}{p^+}\;k^{\,c}\,k^{+\,(2-N-c)}\,l^{\,(3-a-c)}\,l^{+\,(a+c-1)}\nn\\
&+&\,C_3\;q^{\,a}\,q^{+\,(N-a)}\;p^{\,(N-a)}\,p^{+\,(a-1)}\, k^{\,c}\,k^{+\,(2-N-c)}\,l^{\,(3-N-c)}\,l^{+\,(c-1)} \\
&+&\,C_4\;q^{\,(3-a-c)}\,q^{+\,(a+c-1)}\;\frac{1}{p^{+}}\; k^{\,a}\,k^{+\,(2-N-a)}\,l^{\,c}\,l^{+\,(N-1-c)}\bigg\} \,\phi\,\bar{A}^i\,\bar{A}^j\,\bar{A}^k + c.c. \,. \nn
\eea
We write the above expression in terms of spinor helicity variables using the binomial expansion. 
\bea
V^{\beta} &=& \sum_{N=0}^{3}\,(-1)^N \;\Bigg\{\,C_1\; \Big\{\langle l p \rangle ^N\, \langle lk \rangle ^{3-N}\, + \langle pk  \rangle ^N\, \langle pl \rangle ^{3-N}+ \langle kl \rangle ^N\, \langle kp \rangle ^{3-N} \Big\}+ C_2 \;\Big\{ \langle l q \rangle ^N\, \langle lk \rangle ^{3-N}  \nn\\
&& - \;\;\langle kq \rangle ^N\, \langle kl \rangle ^{3-N}- \langle l q \rangle ^N\, \langle lp \rangle ^{3-N}+ \langle pq \rangle ^N\, \langle pl \rangle ^{3-N}+ \langle kq \rangle ^N\, \langle kp \rangle ^{3-N}- \langle pq \rangle ^N\, \langle pk \rangle ^{3-N}\Big\} \nn\\
&& +\;\; C_3 \; \Big\{ \langle pq \rangle ^N\, \langle lk \rangle ^{3-N} + \langle kq \rangle ^N\, \langle pl \rangle ^{3-N}+ \langle lq \rangle ^N\, \langle kp \rangle ^{3-N}  \Big\} \nn\\
&&+\;\;  C_4 \;\Big\{ \langle lq \rangle ^N\, \langle pq \rangle ^{3-N}+\langle pq \rangle ^N\, \langle kq \rangle ^{3-N}+ \langle kq \rangle ^N\, \langle lq \rangle ^{3-N} \Big\}\; \Bigg\}\,.
\eea
\ndt Using momentum conservation and schouten identity the cofficients are found to be $C_1 = -\frac{3}{4}, C_2 = \frac{2}{4}, C_3 = \frac{1}{4}$ and $C_4= \frac{1}{4}$. We obtain the compact form of the quartic vertex
\bea
\label{Vp}
V^{\beta} = \langle pk \rangle \langle kl \rangle \langle l p \rangle \,.
\eea
The vertex (\ref{Vp}) matches with the known result in the literature \cite{Neill:2009mz}. The construction for the case $(0+++)$ follows in a similar manner. The vertices with helicity configurations $(0++-)\,,\,(0--+)$ contain mixed transverse derivatives. For such type of vertices, the spin transformations in both $j^-\,,\,\bar{j}^-$ are non-trivial. Hence, these cannot be uniquely fixed by the algebra. Moreover, at this order, these vertices are proportional to the free equations of motion and hence can be removed by suitable field redefinition.
\vskip 0.3cm
\section{Five-point NMHV amplitude for $R^3$ operator}
The five-point NMHV amplitude can be constructed using the inverse soft method as shown below
\bea
\label{5N}
M_5^{R_+}(1^-,2^-,3^-,4^-,5^+)&=&S(1,5^+,4)M_4(1'^-,2^{-},3^{-},4'^{-})\,+\, S(1,5^+,3)M_4(1'^-,2^{-},3'^{-},4^{-})\nn\\
&+&S(1,5^+,2)M_4(1'^-,2'^{-},3^{-},4^{-})\,+\,S(5,1^-,2)M_4(2'^{-},3^{-},4^{-},5'^+)\nn\\
&+& S(5,1^-,3)M_4(2^{-},3'^{-},4^{-},5'^+) \,+\, S(5,1^-,4)M_4(2^{-},3^{-},4'^{-},5'^+) \nn\\
&+& \left[ S(5,4^-,3)- S(5,4^-,3'') \right] \,S(1,5'^+,3')\, M_3(1^-,2^{-},3'^{-}) \nn\\
&+&\left[ S(5,4^-,2)- S(5,4^-,2'') \right] \,S(1,5'^+,2')\, M_3(1^-,2'^{-},3^{-})\nn\\
&+&\left[ S(5,2^-,3)- S(5,2^-,3'') \right]\,S(1,5'^+,3')\, M_3(1^-,3'^{-},4^{-}) \,,
\eea
where the prime and the double prime indicate appropriate momentum deformations as explained below. \\

\ndt The five-point NMHV amplitude $M_5^{R_+}(1^-,2^-,3^-,4^-,5^+)$ receives contributions from three classes of diagrams. The diagrams are classified based on the seed amplitude and the factorization channel. We first start with $M_4^{R_+}(1^-,2^-,3^-,4^-)$ as a seed amplitude \cite{Broedel:2012rc,Cohen:2010mi}
\bea
M_{4}^{R_+}(1^-,2^-,3^-,4^-)  = 10 \,{stu}\, \frac{\langle 12 \rangle \langle 23 \rangle \langle 34 \rangle \langle 41 \rangle  }{[12][23][34][41]}.
\eea
The soft factor $S$ for graviton leg $5$ of positive helicity is \cite{Boucher-Veronneau:2011rwd}
\bea
S(1,5^+,i) = \sum_{i=2}^4\,\frac{[5i]\, \langle 1i \rangle^2}{\langle 5i \rangle \,\langle 15 \rangle^2}  \,.
 \eea
The first class of diagram has three terms which are related by the interchange of gravitons $2,3,4$. We evaluate one of the terms
\bea
D_1 &=& \frac{[54]\, \langle 14 \rangle^2}{\langle 54 \rangle \,\langle 15 \rangle^2}  \times 10 \, \langle 12 \rangle^2  \langle 23 \rangle  \langle 34 \rangle  \langle 41 \rangle^2  \langle 13 \rangle\, \frac{[31']}{[23][34']} \,,\nn\\
&=& 10\, \frac{ \langle 12 \rangle \langle 23 \rangle \langle 34 \rangle \langle 13 \rangle \langle 24 \rangle  \langle 41\rangle^4\, [54]}{ \langle 54 \rangle \langle 15 \rangle^2 \, [23]} \,,
\eea
where the prime indicates shifts defined in (\ref{Gs}). We now compute the second class of diagrams with  $M_4(2^-,3^-,4^-,5^+)$ as the seed amplitude (\ref{M4}). In this class, we find
\bea
D_2 = \frac{ \langle 12 \rangle \langle 43 \rangle^5 [54]^3 [35]^3}{[12][32][43][42][51]^2}\,,
\eea
plus two more contributions related by the exchange of gravitons $2,3,4$. In the third class of diagrams, the soft gravitons are added to the three-point amplitude with suitable momentum deformations \cite{Boucher-Veronneau:2011rwd}
\bea
\label{D3}
D_3 &=& \left[ S(5,4^-,3)- S(5,4^-,3'') \right] \,S(1,5'^+,3')\, M_3(1^-,2^{-},3'^{-})\,, \nn\\
&=& \left( \frac{\langle 43 \rangle \, [53]^2}{[43]\,[54]^2} -  \frac{\langle 43'' \rangle \, [53'']^2}{[43'']\,[54]^2} \right) \frac{[53] \,\langle 13' \rangle^2}{\langle 5'3' \rangle\langle 15' \rangle^2}\, M_3(1^-,2^{-},3'^{-})\,.
\eea
The double primes indicate deformations with respect to the first added particle $5^+$ and then the second added particle $4^-$. The prime indicate the shift of the angle bracket due to the last added particle $4^-$. The prime and double prime shifts are defined as
\bea
&&|3'\rangle = \frac{(3 + 4)|5]}{[35]}\,, \;\;\; |5'\rangle = \frac{(5+4)|3]}{[53]}\,, \nn\\
&& |3''] = \frac{(3+4+5) |1 \rangle [53]}{[5|\,3+4\,|1 \rangle} \,,\;\;\;\;  \langle 3''| = \frac{[5| (3+4 )}{[53]}\,.
\eea
 Plugging this back into (\ref{D3}), we obtain
\bea
D_3 = \frac{ \langle 12 \rangle^5 \langle 43 \rangle [15]^2 [52]^4 }{[21][23][35][54][43][24]}\,.
\eea
The other two terms in this class of diagram can be obtained by the interchange of gravitons $2,3,4$. \\

\ndt Summing the contributions from all diagrams, the five-point NMHV graviton amplitude for the $R^3$ operator is
\bea
M_5^{R_+}(1^-,2^-,3^-,4^-,5^+)&=& D_1 + D_1(4 \leftrightarrow 3) + D_1(4 \leftrightarrow 2) \nn\\
 &+& D_2 +D_2 (2 \leftrightarrow 3)+ D_2 (2 \leftrightarrow 4) \nn\\
&+& D_3 + D_3 (3 \leftrightarrow 2) + D_3 (4 \leftrightarrow 2)\,.
\eea


\end{document}